\documentclass[]{IEEEtran}

\IEEEoverridecommandlockouts

\input epsf
\usepackage{graphicx}
\usepackage{url}

\usepackage{amsmath}    

\usepackage{xcolor}   

\DeclareRobustCommand*{\IEEEauthorrefmark}[1]{%
  \raisebox{0pt}[0pt][0pt]{\textsuperscript{\footnotesize\ensuremath{#1}}}}

\begin{document}

\title{\LARGE Electrical drive of a Josephson junction array using a cryogenic BiCMOS pulse generator}


\author{\IEEEauthorblockN{
Yerzhan Kudabay,\IEEEauthorrefmark{1},
Oliver Kieler\IEEEauthorrefmark{2},
Michael Starkloff\IEEEauthorrefmark{3},
Marco Schubert\IEEEauthorrefmark{3},
Michael Haas\IEEEauthorrefmark{2},
Johannes Kohlmann\IEEEauthorrefmark{2},
Vadim Issakov\IEEEauthorrefmark{1}, and
Mark Bieler\IEEEauthorrefmark{2}}\IEEEauthorrefmark{*} \\
\IEEEauthorblockA{\IEEEauthorrefmark{1}Institute for CMOS Design, Braunschweig University of Technology, Braunschweig, Germany}\\
\IEEEauthorblockA{\IEEEauthorrefmark{2}Physikalisch-Technische Bundesanstalt, Bundesallee 100, 38116 Braunschweig, Germany}\\
\IEEEauthorblockA{\IEEEauthorrefmark{3}Supracon AG, Jena, Germany}\\
\IEEEauthorblockN{\IEEEauthorrefmark{*}mark.bieler@ptb.de}
}


\maketitle
\IEEEpubidadjcol
\begin{abstract}
We have combined a cryogenic BiCMOS integrated circuit with a superconducting Josephson junction array. The BiCMOS circuit serves as an on-chip pulse generator delivering return-to-zero pulses with pulse frequencies of up to 15\,GHz. With this microwave drive, we observe wide and flat Shapiro steps in the voltage-versus-current characteristics of the Josephson junction array. To the best of our knowledge, this is the first integration of a Josephson junction array with a cryogenic BiCMOS chip. Our results pave the way to a hybrid and fully integrated Josephson arbitrary waveform synthesizer with possible applications in quantum electronics and quantum voltage metrology. 
\end{abstract}


\pagenumbering{gobble}

\section{Introduction}

The Josephson Arbitrary Waveform Synthesizer (JAWS) represents the latest generation of AC voltage standards, delivering an excellent signal-to-noise ratio (SNR) for waveforms from DC up to the MHz range \cite{benz2024, kiel2024}.  Being dependent on the number of Josephson junctions, the SNR may be as low as -120\,dBc or even -140\,dBc \cite{kiel2024, behr2015}. Typically, the Josephson junction array (JJA) of the JAWS circuit is driven by expensive and bulky (19-inch wide) pulse pattern generators held at room-temperature (RT). 

In this work, we demonstrate the feasibility of combining a JJA with a purpose-designed cryogenic BiCMOS pulse generator directly at a temperature of approximately 4~K. In the following the BiCMOS pulse generator, with an area of a few mm$^2$, is referred to as  serializer. We experimentally demonstrate the serializer operation up to 15\,GHz pulse frequency, driving the JJA, consisting of non-hysteretic Nb/NbSi$_x$/Nb Josephson junctions, and leading to wide and flat Shapiro steps in the JJA's voltage-versus-current characteristics. 
Our results denote a first important step towards a fully integrated hybrid JAWS circuit, reducing system complexity, enhancing scalability, and minimizing total power consumption for applications in voltage metrology and cryogenic quantum technology.

\section{Serializer circuit design and characterization}

The concept for our hybrid integration of the JJA and the BiCMOS chip comprises two stages. The first stage, located at RT, contains low-frequency control electronics. The second stage, located at 4\,K, contains the cryogenic serializer integrated circuit. This chip occupies an area of 2 x 2.5 mm$^2$ and was fabricated using Infineon’s 0.13\,µm SiGe B11HFC BiCMOS process. For the work presented in this contribution, only the high-speed serializer, modulator, and clock-distribution network are integrated on the BiCMOS chip allowing for the generation of equidistant pulses. The bit-pattern generation, allowing for arbitrary bit sequences, will be integrated at a later stage.  

\begin{figure}[h!]
    \centering
	\vspace{-0.3 cm}
	\includegraphics[width=0.85\columnwidth]{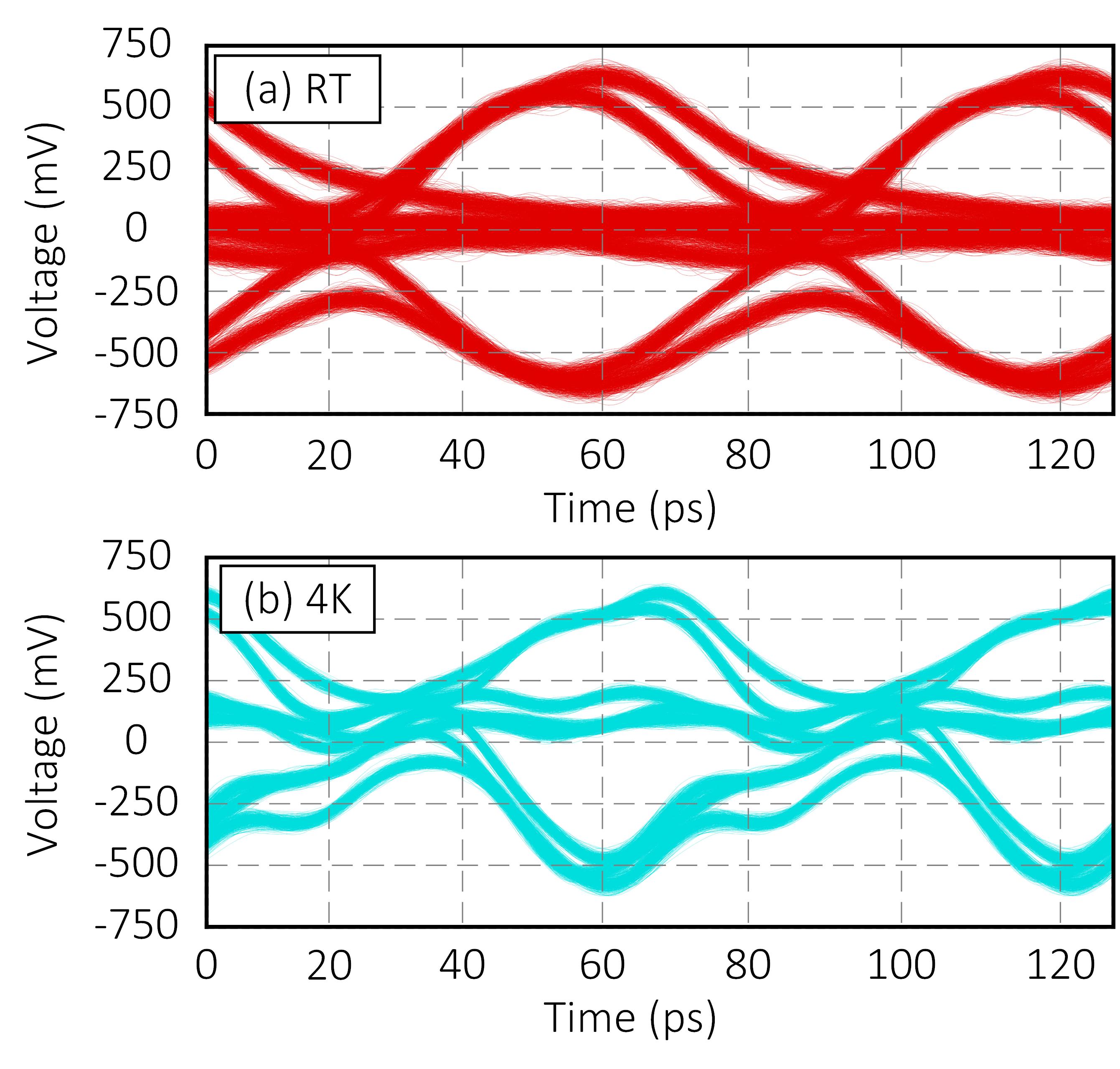}
	\vspace{-.5 cm}
	\caption{\label{fig:eye} Measured eye diagrams of the serializer with a pulse frequency of 15\,GHz (30\,Gb/s bit sequence: -10-10+10-1000+100000) at RT (a) and at approximately 4\,K (b).}
\end{figure}

The cryogenic serializer has previously been introduced in Refs.~\cite{kuda2025,kuda2024}. Yet, no experimental results combining the serializer and a JJA have been published to date. The serializer chip was characterized under multiple operating conditions. Output eye diagrams with 15\,GHz pulse frequency (corresponding to 30\,Gb/s data rate, since two data bits, "1" and "0", form one return-to-zero pulse) and measured at RT and approximately 4\,K are shown in Fig.~\ref{fig:eye}(a) and Fig.~\ref{fig:eye}(b), respectively. 
A noticeable change of the eye diagram is observed at cryogenic temperatures. This behavior is attributed to temperature-dependent changes in the dielectric properties of substrate and interconnect materials, leading to overshoot, ringing, and steeper signal transitions that increase inter-symbol interference. For the measured eye diagrams shown in Fig.~\ref{fig:eye} the serializer's power consumption was 306\,mW at RT and 436\,mW at approximately 4\,K with less than 2\% variation of power consumption upon change of the bit sequence.

\section{Combined serializer-JJA measurements}

The two JJAs used in this work comprise 150 and 7494 Josephson junctions, embedded in the center line of a coplanar waveguide, which was matched to 50\,$\Omega$ impedance. The JJA chips have dimensions of 10 × 10 mm$^2$ and were fabricated in PTB’s clean-room. More details about fabrication of JJAs used in this work can be found in Refs.~\cite{kiel2024, kiel2015}. The JJAs exhibit a critical current of $I_\text{c} = 4.3$\,mA for the smaller array and $I_\text{c} = 5.8$\,mA for the larger array, a normal-state resistance of $R_\text{n} = 3.1$\,m$\Omega$, a characteristic voltage of $V_\text{c} = 13.5$\,µV, and a corresponding characteristic frequency of $f_\text{c} = 6.5$\,GHz. 

The performance of the JJAs and the BiCMOS serializer was evaluated under cryogenic conditions using both a liquid Helium (LHe) Dewar and a pulse-tube cryocooler made by Supracon, thereby demonstrating the robustness and reproducibility of the joint JJA and serializer measurements. 
For the LHe measurements, the JJA consisting of 150 Josephson junctions and the BiCMOS serializer were mounted on two separate printed circuit boards, each equipped with two RF connectors. The boards were attached to a cryogenic probe stick and interconnected by a 10\,cm long coaxial cable to transmit the serializer output signal to the JJA. The remaining connectors were used to supply the clock signal to the serializer and to provide mechanical fixation. Electrical isolation was achieved using inner and outer DC blocks.
For the cryocooler-based measurements, the serializer and the JJA comprising 7494 Josephson junctions were mounted on a dedicated copper carrier. Both chips were connected via wire bonds and the copper carrier was subsequently installed inside the cryocooler. The serializer PCB was equipped with an RF connector to provide the high-frequency clock signal. Electrical isolation of the JJA was ensured by inner and outer DC blocks integrated on the JJA chip. 

\begin{figure}[h]
    \centering
	\includegraphics[width=0.82\columnwidth]{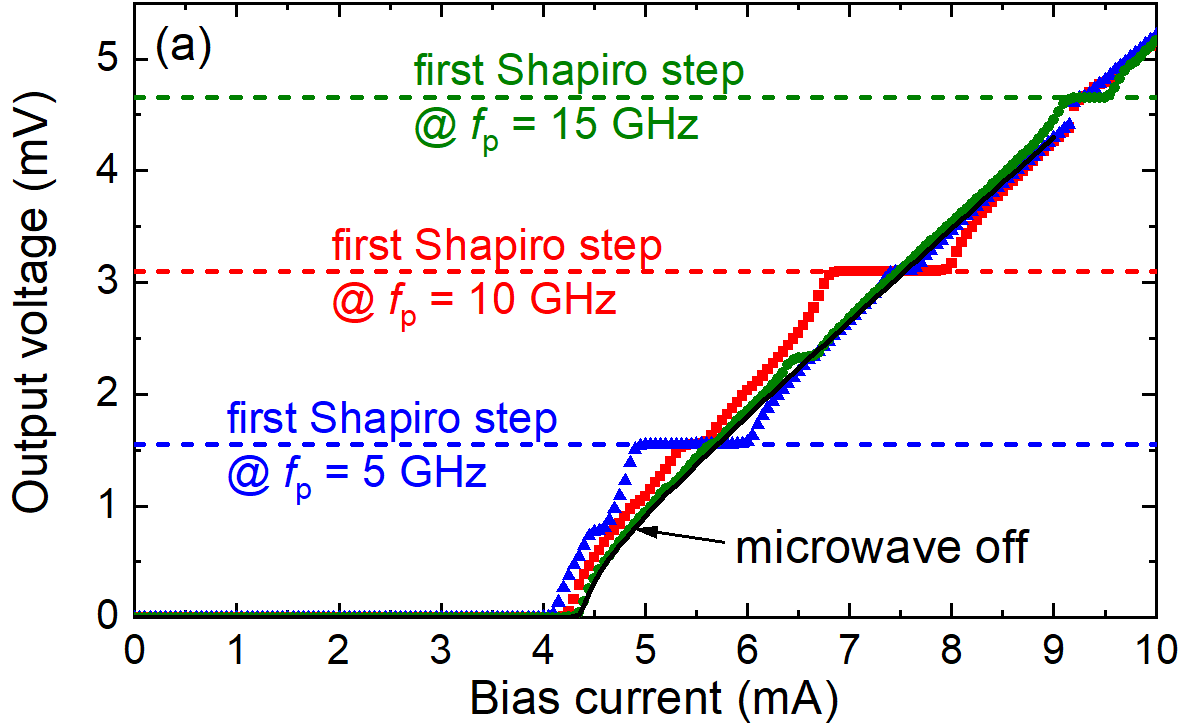}
    \includegraphics[width=0.82\columnwidth]{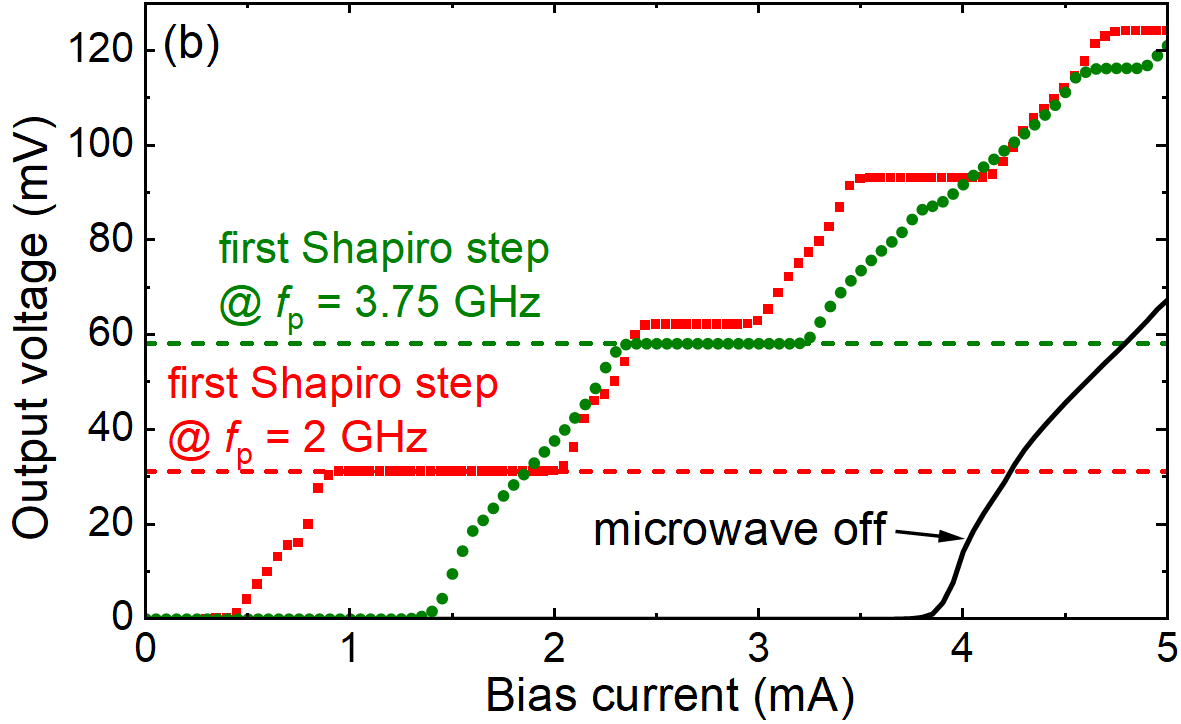}
	\vspace{-0.2 cm}
	\caption{\label{fig:Shapiro} Voltage-versus-current characteristics of JJAs being driven by the cryogenic serializer in liquid Helium at 4.2\,K (a) and in a cryocooler at 4.9\,K (b). The bit sequences were +10+10+10+10... in (a) and +1000-1000+1000\linebreak -1000... in (b) and the data rate has been adjusted to obtain different $f_\text{p}$.}
    \vspace{-0.3 cm}
\end{figure}

The results of the cryogenic measurements are summarized in Fig.~\ref{fig:Shapiro}. Pronounced Shapiro steps are observed for the JJA with 150 junctions in LHe at pulse frequencies $f_\text{p}$ of 5\,GHz, 10\,GHz, and 15\,GHz, as well as for the JJA with 7494 junctions in the cryocooler at 2\,GHz and 3.75\,GHz. The presence of wide and well-defined Shapiro steps confirms that the serializer output signal is sufficiently preserved when transferred to the JJA chips under cryogenic conditions with different measurement setups, different JJAs and at different frequencies. Moreover, Fig.~\ref{fig:Shapiro}(b) demonstrates that for low pulse frequencies we even observe clear Shapiro steps up to the fourth order. 

\section{Conclusion}
In conclusion, we have successfully integrated a high-speed cryogenic BiCMOS serializer, achieving pulse frequencies (data rates) up to 15\,GHz (30\,Gb/s), with  superconducting JJAs~\cite{kuda2025b}. Wide and flat Shapiro steps were observed. To the best of our knowledge, no system combining a high-speed cryogenic driver with JAWS or JJA has been reported in the literature so far with this level of integration. Our results mark a significant step towards the co-integration of classical, CMOS-based control electronics with superconducting quantum circuits with the aim to realize a scalable, quantum-accurate, and ultra-low-noise signal generation platform at cryogenic temperatures.

\section*{Acknowledgment}
This work was funded by the German Federal Ministry of Research, Technology and Space (BMFTR) within the Funding Programs "Quantum Technologies - from basis research to market" (contract number: 13N15934) and "Quantum Systems" (contract number: 13N17247).



%

\end{document}